# Significance of Coupling and Cohesion on Design Quality

Poornima U. S.[1], Suma. V.[2]

[1,2]Research and Industry Incubation Centre, Dayananda Sagar Institutions
[1]Raja Reddy Institute of Technology, Bangalore, India.
[1]uspaims@gmail.com, [2]sumadsce@gmail.com

*Abstract*—In recent years, the complexity of the software is increasing due to automation of every segment of application. Software is nowhere remained as one-time development product since its architectural dimension is increasing with addition of new requirements over a short duration. Object Oriented Development (OOD) methodology is a popular development approach for such systems which perceives and models the requirements as real world entities. Classes and Objects logically represent the entities in the solution space and quality of the software is directly depending on the design quality of these logical entities. Cohesion and Coupling (C&C) are two major design decisive factors in OOD which impacts the design of a class and dependency between them in complex software. It is also most significant to measure C&C for software to control the complexity level as requirements increases. Several metrics are in practice to quantify C&C which plays a major role in measuring the design quality. The software industries are focusing on increasing and measuring the quality of the product through quality design to continue their market image in the competitive world. As a part of our research, this paper highlights on the impact of C&C on design quality of a complex system and its measures to quantify the overall quality of software.

## I. INTRODUCTION

Object Orientation is a popular development suit for the application domain with scalable requirements over time. The customer needs are collected as entities in a realistic way and modelled as logical types for a programming language. The success of such projects with huge set of requirements depends on the quality of development process from requirements collection to product testing. The development method that is being followed in the industry to carry out development process as System Development Life Cycle (SDLC) is different. The workflow process which industry adapts includes all the steps of SDLC for each project module under development. Among all phases, the design phase is a crucial phase which entitles the system architecture and logical interface as work products. At design phase, the requirements in the problem statement are classified as logical types like classes, interfaces, delegates, structures and enumerations such that the solution space site the representative of real word objects.

The design of a complex system needs to be flexible enough to incorporate the changes in future. This flexibility of design is based on structure of the logical types and their interdependency. Good design supports tight interrelationship within the types, namely cohesion and less dependency between the modules, namely coupling. The solution space of the problem statement holds several types of cohesion and coupling which is a vital part of the system design operates as a decisive factor for a good design. Framing cohesion and coupling is thus very challenging for the designer since design fault increases with the system complexity [1]. Hence, industries are adapting several design patterns, architectures, procedures and best practices to improve the overall quality of the system design [2].

Besides, several qualitative and quantitative metrics are in practice to measure the quality of work process and the final product. Such quantitative metrics are more applied in the industry rather than the qualitative metrics which needs to be made handy for the design quality access team. The design quality metrics for cohesion and coupling for OOD plays a prime role in qualifying the overall design. However, it has been remaining as an open issue for the researchers to work on design quality metrics for better system quality.

## II. OBJECT ORIENTED SYSTEM MODELLING

Modelling is a primary phase of software development which represents the real world entities as logical units. There is a clear separation between the attributes of an entity which characterises it during system modelling. Abstraction is the basic principle of modelling which creates a view of an entity relevant enough to the present scenario. In the later stage of modelling, abstraction unfolds more on the entity which provides different level of views, revealing more on the module/system under development. Hence, a model provides a detail pictorial description of problem statement in hand [3].

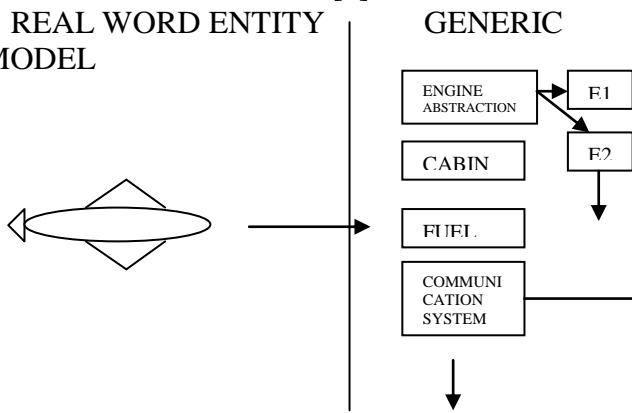

Fig.1. Modelling of real world entities. View 1 ………View N

The process of modelling at different level provides different views. Thus, abstraction open ups various views of the upcoming system. The ability to frame the abstraction contributes to quality design in terms of choosing the modules and thereby modelling the system at different levels of complexity.

System model can be viewed to be static or dynamic. Static model represents the system architecture which includes classes and modules diagrams. The interaction between the objects is depicted through activity diagrams and interaction diagrams showing the dynamic nature of the project in hand [4].

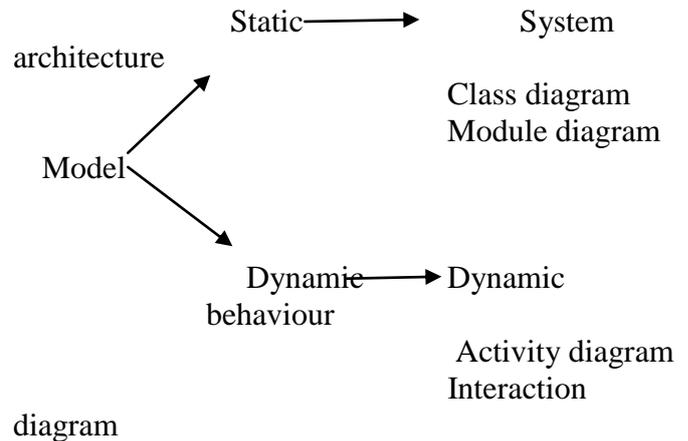

Fig.1. Modelling in Object Oriented System.

The static and dynamic modelling of the project contributes to overall quality of the software design. In static design, the data and operational attributes in the logical types are expected to be much interrelated to each other to support separation of concern. This cohesive nature thereby constructs the logical types as individual entities which supports scalability and maintainability of the system at ease. Besides, the logical dependency either on data or services, which is represented by dynamic model, expected to be minimised to avoid cross coupling which increases the design complexity.

## III. COUPLING AND COHESION AS DESIGN QUALITY DESSISIVE PARAMETERS

The design of an application software with huge set of requirements obviously imbibes the complexity. Such applications are more likely to augment in future in terms of services as requirement grows. Hence, the design architecture of such systems is flexible enough to incorporate the changes even after the deployment as a part of enhancement. This could be possible when the application design is made more flexible and the relationship between the logical types is under control. This is achieved by framing highly cohesive and low dependent modules [5][6]. The design team needs to have a measure of these design qualities to control the complexity during design process. Several metrics are in practice to measure the work products of SDLC [7][8][9]. Cohesion and coupling metrics available in the literature do the assessment on the some parameters of cohesion and coupling [10][11][12][13][14].

*A. Static / Class level coupling and its Metrics*

Static coupling represents the permanent code binding between the modules. In framing the solution space, the individual logical types like classes, structures, enumerations, delegates., need to be interconnected within or between the modules. The flexibility of the architecture depends on the degree of the connectivity between them. Static coupling appears as either association between two different classes or as an inheritance lattice. Software developers are measuring the association dependency between the modules using two coupling metrics, Efferent and Afferent coupling.

*1) Efferent Coupling:* It measures the degree of dependency of a class on outside packages classes. Dependency is much in such coupling as modification in the parent module has ripple effect on the dependents.

$$Ce = \sum_{i=0}^{n}(C))$$
Eq.1.

Where C is number of outside classes.
Low value of Ce is optimal for good design.

*2) Afferent Coupling :* It is a measure of number of package classes being used by other classes. Thus, it is a parent classes upon which other classes are dependent. It needs much attention when modifications are required.

$$Ca = \sum_{i=0}^{n}(C))$$
Eq.2.

Where C is number of classes being dependent on.
Low value of Ca is optimal for good design.

*3) Depth of Inheritance Tree:* It measures the vertical growth of a class. It supports reusability, however, the complexity directly depends on the depth of the inheritance tree since it is difficult to access the behaviour of the end class in the tree.

$$DIT = \sum_{k=0}^{n}\binom{n}{k}C$$
Eq.3.

Where C is number of classes in a lattice.
The optimal value of DIT is yet a challenge for the design group.

*B. Dynamic / Object level Coupling and Metrics*

Dynamic coupling is the dependency between the classes at runtime by passing the arguments through class methods. This data coupling is a crucial bondage between the classes which shares the services by passing the objects as arguments.

*1) Coupling Between Objects:* It measures the interdependency between the classes. It reflects the sharing of services between the classes in a solution space. This is achieved by using an object as an argument of other class in a existing class member to import the required service. Since it exhibits a dependency, design group requires a statistical value to predict the design complexity.

$$CBO = \sum_{k=1}^{n}\left(C(M(O)\,_{k}^{n})\right)$$
Eq.4.

Where M(O) represents number of services been called.

*2) Response For a Class:* It counts the number of methods invoked in response to a message sent by an object. The methods are present either inside a class or outside.

$$RFC = \sum_{k=0}^{n}\left(M\,_{k}^{n}\right)$$
Eq.5.

Where M is count for methods invoked.

*C. Cohesion as complexity reducing factor*

Cohesion represents the relevancy of the data and the corresponding methods in a class where it is defined. Abstraction process identifies these data set and methods which it uses and prepares a logical model of a problem statement in hand. However, the relevancy is much important to reduce the class complexity. If not, the class is decoupled to reduce the complexity which also supports the separation of concern. Several metrics are proposed in the literature to measure cohesion of a class type.

*1) Lack of COhesion in Methods :* It measures the quality of a class in a solution domain. Cohesion refers the degree of interconnectivity between attributes of a class. A class is cohesive if it cannot be further divided in to subclasses. LCOM measures the method's behaviour and its relevance where it is defined. Pair of methods using data object proves the cohesiveness where as the methods not participating in data access makes it less cohesive. The paired and unpaired methods of classes participating in data access to prove cohesiveness could be calculated by the intersection of object sets related to them. C is a class and M1,M2...Mn are its methods using set of class instances. I1={a,b,c,d}, I2={a,b,c} and I3={x,y,z} are set of instances used by the methods M1,M2 and M3 respectively. If intersection of object set is non-empty then the method using them is cohesive and their relevance in the class is proved. i.e. I1 ∩ I2 = {a, b, c} means M1 and M2 are cohesive. But intersection of I1, I3 and I2, I3 is empty set. Thus the relevancy of $M_3$ is less the class and it is decoupled with the method $M_3$. High count in LCOM shows less cohesiveness and class need to be divided to subclasses.

$LCOM = \{I1 \cap I3\} = \{\varnothing\}$
Eq.6.

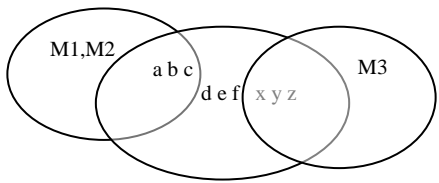

*2) LCOM2, LCOM3, LCOM4*

Different variations of LCOM have been practiced in the literature to increase the cohesiveness of the class attributes.

$LCOM_2 = P-Q$
        Eq.7.

Where P= Number of pairs of methods that do not share the attributes and Q= Number of pairs of methods that share attributes.

$LCOM_3$= Number of connected components in the graph that represents each method as a node and the sharing of at least one attribute as an edge.

TABLE I
COUPLING AND COHESION METRICS

| Sl. No | Coupling and Cohesion Metrics | | |
|---|---|---|---|
| | Metric | Formula | Description |
| 1 | Efferent Coupling | $Ce = \sum_{i=0}^{n}(C))$ | Measures dependency on outside classes |
| 2 | Afferent Coupling | $Ca = i = 0i = n(C))$ | Measures classes dependent on a class |
| 3 | Depth of Inheritance | $DIT = \sum_{k=0}^{n}\binom{n}{k}C$ | Counts number of classes in a inheritance lattice |
| 4 | Coupling Between Objects | $CBO = \sum_{k=1}^{n}(C(M($ | Measures coupling between classes through objects |
| 5 | Response For a Class | $RFC = \sum_{k=0}^{n}(M\binom{n}{k})$ | Measures the number of methods responded for an object message |
| 6 | Lack of Cohesion in Methods | $LCOM = \{I1 \cap I3\} = \{\varnothing\}$ | Counts the methods not accessing the data |
| 7 | Lack of Cohesion in Methods version 1 | $LCOM_2 = P-Q$ | Difference of paired and unpaired methods |

IV. CORRELATION BETWEEN COHESION AND COUPLING IN MEASURING COMPLEXITY

Degree of cohesion has an impact on coupling as well as complexity of a class. The binding of data attribute and services plays an important role in deciding the quality of abstraction as well as the class complexity. To achieve cohesion, in OO systems, the methods are segregated as selectors and modifiers depending on their accessibility to the data set. Thus the separation of concern decouples the class into a new class with set of relevant methods [15].

A class is highly cohesive when attributes and methods are sufficient and optimal. Sufficiency reduces the dependency on other class services and optimality deduces the complexity. Thus the coupling is reduced with high cohesion. However, the complexity of classes increases even with relevant attributes. At a breakeven point, the class again needs to be decoupled.

## V. EMPIRICAL STUDY ON C & C METRICS

This research focuses upon evaluating different design quality metrics on open source java programs [16][17[18]. The sampled programs comprise of class structures and relationships in order to evaluate the metrics. An open source tool is used to evaluate the coupling and cohesion metrics. The coupling metrics such as DIT, CBO, CA, CE, RFC are expected to be inversely proportional to the value of LCOM to prove a good class design. The low value of LCOM reflects high cohesive class which encourages more decoupling of classes. This assessment is taken as instance of our research and further study is focusing on evaluating the metrics on more complex programs and their intercreativity with parameters of research interest.

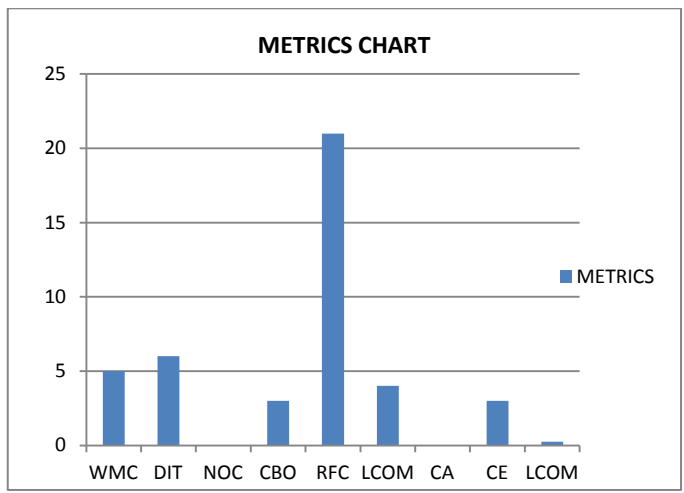

Fig.2 An Object Oriented metric assessment on sample Open Source Java program using Metric assessment tool.

## VI. CONCLUSION

Producing the quality is one of the aims of all industries irrespective of the complexity of the problem statement in hand. However, the development methodology adapted by the industry plays an important role in attaining the quality product. It is very necessary to equip the development team to measure the quality of the work products after each phase of software development process. Hence the quality of the process reflects the quality of the end product. Measures are required to quantify the quality of phases which serves as indicators to the rest of the development process. In complex systems, the design architectures need to be flexible and the complexity needs to be measurable in terms of Cohesion and Coupling existing in the system design. The design quality metrics available are able to assess them to some extent, however much detailed evaluation of metrics is yet to be explored.

## ACKNOWLEDGMENT

The authors would like to sincerely acknowledge all the industry personnel for their valuable suggestions, help and guidance in carrying out this part of research. The complete work is undertaken under the framework of Non Disclosure Agreement.